\documentclass[aps,
groupedaddress,showpacs,nofootinbib,prl,twocolumn]{revtex4}
\usepackage{amssymb,amsmath}
\usepackage{graphics}
\usepackage[english]{babel}
\usepackage{graphicx}
\usepackage{dcolumn}
\usepackage{bm}

\def\ins#1{{\it #1}}

\def\meq#1{}
\def\meq#1{}

\def\ins#1{}

\def\comment#1{}

\def\mn#1{\marginpar[]{\scriptsize#1}}
\def\mn#1{}

\usepackage{ulem}
\usepackage{cancel}
\usepackage{color}
\def\rr#1{\textcolor{red}{#1}}

\def\mn#1{\marginpar[\tiny{\rr{#1}}]{\tiny{\rr{#1}}}}

\def\rr#1{}

\usepackage{hyperref}

\def\comment#1{}

\newcommand{\be}{\begin{equation}}\newcommand{\ee}{\end{equation}}
\newcommand{\bea}{\begin{eqnarray}}\newcommand{\eea}{\end{eqnarray}}
\newcommand{\beaa}{\begin{eqnarray}}\newcommand{\eeaa}{\end{eqnarray}}
\newcommand{\ba}{\begin{array}}\newcommand{\ea}{\end{array}}
\newcommand{\bit}{\begin{itemize}}\newcommand{\eit}{\end{itemize}}
\newcommand{\ben}{\begin{enumerate}}\newcommand{\een}{\end{enumerate}}
									\def\be{\begin{equation}}
\def\ee{\end{equation}}				\def\bea{\begin{eqnarray}}				\def\eea{\end{eqnarray}}
\def\bear{\begin{array}}				\def\eear{\end{array}}

						\def\A{\Alpha}

												\def\x5{x^{5}}

\begin{document}

\title{The Purely Geometric Part of ``Dark Matter''--\\A Fresh
  Playground for ``String Theory''
}

\author{Hagen Kleinert}
\email{h.k@fu-berlin.de}


\affiliation{Institut f{\"u}r Theoretische Physik, Freie Universit\"at Berlin, 14195 Berlin, Germany}
\affiliation{ICRANeT Piazzale della Repubblica, 10 -65122, Pescara, Italy}


\vspace{2mm}

\begin{abstract}

We argue that part of ``dark matter'' is not made of matter, but 
of the singular world-surfaces in
the solutions of
{\it Einstein's vacuum field equation\/}
 $G_{\mu\nu}=0$.
Their Einstein-Hilbert action 
governs also their  quantum fluctuations.
It coincides with the 
action of closed bosonic ``strings'' 
in four spacetime dimensions, 
which appear here in a
new physical context.
Thus, part of 
dark matter is of a purely geometric nature, and
its quantum physics is governed by 
the same string theory, whose 
massless spin-$2$
particles interact like
the quanta of Einstein's theory.


\end{abstract}


\maketitle



Dark matter was postulated by
F. Zwicky in 1933 to explain the ``missing  mass'' 
in the orbital velocities of galaxies in clusters. 
Later indications came from measurements 
of the orbital motion of
stars inside a galaxy, where a plot of orbital velocities versus distance
from the center was attributed
to
large amounts of
invisible matter.
The Friedmann model of
 the evolution of the universe 
indicates
that dark matter
constitutes roughly 83\% of the
mass energy of the universe, and there are many speculations
as to its composition. 
In this note we want to propose the simplest possible explanation
of a part of it.

As a warm-up, let us remember
that all static electric fields in  nature 
may be considered as originating 
from the nontrivial
solutions of the 
Poisson equation 
for the electric potential $\phi(x)$:
\begin{eqnarray}
\Delta \phi(x)=0.
\label{@}\end{eqnarray}
The simplest of them has the form $e/r$, 
and is
 attributed to a
pointlike electric charges, whose size $e$
can be extracted  
from the pole strength of the singularity. This becomes
visible 
by performing a spatial integral 
over $\Delta \phi(x)$, 
which yields
$-4\pi e$,
 after applying Gauss's integral theorem. 
Hence the right-hand side of the 
Poisson
equation is not strictly zero, 
but should more properly be 
expressed with the help of  a Dirac-delta function 
$\delta^{(3)}({ x})$ as
\begin{eqnarray}
\Delta \phi(x)=-4\pi e\delta^{(3)}({\bf x}).
\label{@}\end{eqnarray}

For celestial objects, the situation is quite similar.
The
Einstein equation in the vacuum,
$G_{\mu\nu}=0$,
possesses simple nontrivial  solutions
in the form of the Schwarzschild metric defined by
\begin{eqnarray}
ds^2&=&g_{\mu\nu}dx^\mu dx^\nu\nonumber
=(1-r_S/r)c^2dt^2-(1-r_S/r)^{-1}dr^2
\\&-&r^2(d\theta
^2+\sin ^2 \theta d\varphi^2),
\label{@}\end{eqnarray} 
with $r_S\equiv 2GM/c^2$ being the Schwarzschild radius,
or its rotating generalization, the Kerr metric.
Also here we may  calculate 
the spacetime integral over the
homogeneous 
Einstein equation, to find a nonzero result, namely  
\begin{eqnarray}
\int d^3 x \,G_{0}{}^{0}= \kappa c M,
\label{@MSSR}\end{eqnarray}
where $\kappa$ is the gravitational constant 
defined in terms of Newton's constant 
$G_{\rm N}$, or the Planck length $l_{\rm P}$, as
\begin{eqnarray}
\kappa\equiv 8\pi l_{\rm P}^2/\hbar=
 8\pi G_{\rm N}  /c^3.
\label{@}\end{eqnarray}
From (\ref{@MSSR}) we identify the mass of the object
as being $M$.

If the mass point moves through spacetime along a trajectory
parametrized by $x^\mu(\tau)$,
it has an energy-momentum tensor 
\begin{eqnarray}
T^{\mu\nu}(y)=M\int_{-\infty}^\infty d\tau
\dot x^\mu(\tau)\dot x^\nu(\tau)
\delta^{(4)}(y-x(\tau)),
\label{@}\end{eqnarray}
where a dot denotes the $\tau$-derivative.
We may integrate the associated 
solution of the homogeneous Einstein equation 
$G_{\mu\nu}=0$ 
over spacetime, and find,
using $\dot x^2=1$, that
that its
Einstein-Hilbert action 
\begin{eqnarray}
{\cal A}_{\rm EH}=-\frac1 {2\kappa}\int d^4 x\sqrt{-g}R,
\label{@EH}\end{eqnarray}
is proportional  
to
the classical action of 
 a point-like particle:
\begin{eqnarray}
{\cal A}_{\rm EH}^{\rm world line}\,\mathop{\propto}\,
-Mc\int ds,
\label{@EHAC1i}\end{eqnarray} 
 proportional  
to
the classical action of 
 a point-like particle.
A slight modification 
of (\ref{@EHAC1i}), that is the same classically, 
but different for fluctuating orbits,
describes also the quantum physics of a spin-$0$ particle \cite{PI}
in a path integral over all orbits.
Thus Einstein's action 
for a singular world line in spacetime
can be used to define also the quantum physics a spin-$0$ point particle.

In addition to pointlike singularities,
the homogeneous Einstein equation will also 
possess singularities on surfaces in spacetime. 
These may be 
parametrized by $x^\mu(\sigma,\tau)$, and their energy-momentum tensor
has the form
\begin{eqnarray}
T^{\mu\nu}(y)\propto\int_{-\infty}^\infty d\sigma d\tau
( \dot x{}^\mu
\dot x^\nu \!-\!
 x'{}^\mu
 x'{}^\nu)
\delta^{(4)}
(y-x(\sigma,\tau)),
\label{@}\end{eqnarray}
where a prime denotes a $\sigma$-derivative.
In the associated Einstein tensor, the 
$\delta$-function on the surface leads to a volume integral 
\cite{REMAA}:
\begin{eqnarray}\!\!\!\!\!\!\!\!\!\!\!\!
\!\!\!\!\!
&&\int d^4x\sqrt{-g}\, 
G_{\mu}{}^{\mu}\!\propto\! 
 \int\! d^2 a
\equiv\!\!\!  \int \!
d\sigma d\tau 
\sqrt{
(\dot xx'{})^2-
\dot x'{}^2x'{}^2
}.
\label{@}\end{eqnarray}
By analogy with the line-like case we 
obtain
for such a singular 
field
an Einstein-Hilbert action (\ref{@EH})
\begin{eqnarray}
{\cal A}_{\rm EH}^{\rm worldsurface}\propto-\frac1 {2\kappa}\int d^2a
=
-\frac{\hbar} {16\pi l_{\rm P}^2}\int d^2a.
\label{@EHAC}\end{eqnarray}
Apart from a numerical proportionality factor of order one,
this is precisely the Nambu-Goto action of 
a  bosonic closed string in four spacetime dimensions:
\begin{eqnarray}
{\cal A}_{\rm NG}=
-\frac{\hbar} {2\pi l_{\rm s}^2}\int d^2a,
\label{@EHAC1}\end{eqnarray}
where $l_{\rm s}$ is the so-called
string length
$l_{\rm s}$, related to the string tension $\alpha'$ by
$l_{\rm s}=\hbar c \sqrt{\alpha'}$.
Note that in contrast to the world lines,
there is no extra mass parameter $M$.

The original string model was proposed to describe
 color-electric flux tubes and their Regge trajectories whose slope
 $dl/dm^2=\alpha'$ lies around $1$ GeV${}^ {-2}$.
However,
since the tubes are really fat objects, as fat as pions,
 only very long flux tubes are approximately line-like.
Short tubes 
degenerate into
spherical ``MIT-bags'' \cite{BAG}.
The flux-tube role of strings
was therefore abandoned, and the action
(\ref{@EHAC1})
was
re-interpreted in a completely different fashion,
 as describing the fundamental particles of nature,
assuming $l_{\rm S}$
to be of the order of $l_{\rm P}$.
Then the spin-$2$ particles 
of (\ref{@EHAC1}) would interact like gravitons
and define Quantum Gravity. 
But also the ensuing ``new string theory'' 
 \cite{ST}
has been criticized 
by many authors
\cite{SCHR}. One of its most embarrassing 
failures is
that it has not  produced 
any experimentally observable results. 
The particle spectra 
of its solutions have not matched 
the existing particle  spectra.
The proposal of this note cures this problem.
If ``strings'' describe 
``dark matter'',
there would be no need to reproduce other observed particle spectra.
Instead,
their celebrated virtue, 
that
their
spin-$2$ quanta 
interact like gravitons,
can be used to fix the 
proportionality factor between the Einstein action 
action
(\ref{@EHAC}),
and the string action 
(\ref{@EHAC1}).

It must be kept in mind that 
just as $-Mc \int ds$ had to be modified for fluctuating paths \cite{PI}, also
the Nambu-Goto action 
(\ref{@EHAC1})
needs a modification,
if the surfaces fluctuate. That was
found  
by Polyakov when 
studying 
the consequences of the conformal symmetry 
the theory.
He replaced
 the action 
(\ref{@EHAC1})
by
a
new action that is equal to 
(\ref{@EHAC1}) at the classical  level, but 
contains in $D\neq 26 $ dimensions another spin-0 field 
with a Liouville action.

Since the singularities of Einstein's fields possess 
only gravitational interactions, 
their identification  
with ``dark matter'' seems very natural.
All visible matter consists of singular solutions 
of the Maxwell equations and  
the field equations
of the standard model.
A grand-canonical ensemble of these and
the smooth wave solutions
of the standard
model 
explain an important part of the 
matter in the Friedmann model of cosmological evolution.

But the main 
contribution 
to the energy comes from the above singularities 
of Einstein's equation.
Soon after the universe was created,
the temperature was so high
that the configurational entropy
of the surfaces overwhelmed
completely the impeding Boltzmann factors.
Spacetime was filled with these surfaces 
in the same way as superfluid helium is filled 
with the world-surface of 
vortex lines. In hot helium, these lie so densely packed that the
superfluid behaves like a normal fluid \cite{GFCM,MVF}.
The Einstein-Hilbert action of such a singularity-filled
turbulent geometry 
behaves like the action
of a grand-canonical ensemble of world surfaces 
of a bosonic closed-string model.
Note that here these are two-dimensional 
objects living in four spacetime dimensions, and 
there is definite need to understand their 
spectrum
by studying the associated Polyakov action,
without circumventing
the accompanying 
Liouville field
by escaping 
into unphysical dimensions

It should be noted
that in the immediate neighborhood of the 
singularities, the curvature 
will be so high, 
that Einstein's linear approximation $-(1/2\kappa) R$ to the
Lagrangian
must break down and will have to be corrected by
some nonlinear function of $R$, that starts out like Einstein's,
but continues differently.
A possible modification has been suggested a decade ago 
\cite{KSchm}, and many other options have been investigated since then
\cite{more}.

After the big bang,
 the universe 
expanded and cooled down, so that
large singular surfaces shrunk by emitting 
gravitational radiation.
Their
 density decreased,
and some phase transition made
the cosmos homogeneous and isotropic on the large scale
\cite{REMA}.
But it remained filled with gravitational radiation 
and small singular
surfaces 
that had shrunk until their sizes
reached
the levels
stabilized b quantum physics, i.e.,
when their fluctuating action
decreased to order $ \hbar$.
The statistical
mechanics 
of this
cosmos 
is the analog of a spacetime
filled with superfluid helium
whose specific heat is governed by 
the zero-mass
phonons and by rotons.
Recall  \cite{LLAN},
that in this way Landau discovered the 
fundamental excitations called rotons,
whose existence he  deduced
from the temperature behavior 
of the specific heat.
In the universe, the role of rotons is played by 
the smallest
surface-like singularities 
of the homogeneous Einstein
equation, whose existence 
we deduce from the 
cosmological requirement of dark matter.

The situation 
can also
be illustrated by a further analogy with a many-body system.
The defects in a crystal
whose ``atoms''
have a lattice spacing $l_{\rm P}
$
simulate precisely 
the
mathematics
of a Riemann-Cartan spacetime,
in which disclinations and dislocations
define 
curvature and torsion \cite{KONDO,GFCM2,MVF}.
Thus we may imagine a model of the universe 
as a ``floppy world crystal'' \cite{OA},
a liquid-crystal-like phase  \cite{ZA}
in which a first melting transition 
has led to
correct  
gravitational $1/r$-interactions
between disclinations.
The initial hot universe was filled with defects---it was a ``world-liquid''.
After cooling down to the present liquid-crystal state,  
there remained plenty of residual defects around,
which form our ``dark matter''.

We know that the cosmos is filled 
with a cosmic microwave background (CMB) of photons
of roughly 2.725 Kelvin,
the remnants of the big bang.
They contribute to the Friedmann equation of motion 
a constant $\Omega_{\rm rad} h^2=(2.47 \pm 0.01) \times 10^{-5}$,
where $h=0.72\pm0.03$ is the Hubble parameter,
defined in terms of the Hubble constant
$H$ by $h\equiv H/(100$ km/Mpc sec). The symbol 
$\Omega $ denotes the energy
density divided by the so-called critical density $\rho_{\rm c}\equiv3 H^2/8\pi
G_{\rm N}=1.88\times 10^{-26}h^2$\,kg/m${}^3$
\cite{NAK}.
The baryon density
contributes
 $\Omega_{\rm rad} h^2=0.0227
\pm 0.0006$, 
or 720 times as much,
whereas
the dark matter
contributes  
 $
\Omega_{\rm dark} 
h^2=0.104\pm 0.006$, 
or 
4210 as much.
\comment{
Thus 
the cosmos 
is also filled with the remnants of the singularities of the
Einstein equation in the turbulent spacetime
of the big bang \cite{REMA}.}
If we assume for a moment that all 
massive strings are frozen out, 
and that only the 
subsequently emitted  
gravitons
form a thermal 
background \cite{REMATB}
then, since the energy 
of massless states is proportional 
to $T^4$, the temperature of this
background would be
$T_{\rm DMB}\approx 4210^{1/4}\approx 8 T_{\rm CMB}\approx 22 $K.
In general we expect the presence of also 
the other singular solutions of Einstein's equation
to change this result.

There is an alternative way
of deriving the above-described properties of the 
fluctuating 
singular surfaces of Einstein's theory.
One may rewrite Einstein's theory as a gauge theory 
 \cite{GFCM2,MVF},
and put it on a spacetime lattice \cite{SSS}. Then 
the singular surfaces are built explicitly from plaquettes,
as in
lattice gauge theories of asymptotically-free nonabelian gauge
theories \cite{WILSON}. 
In the abelian case, the surfaces 
are composed 
as shown in Ref.~\cite{KM},
for the nonablian case, see
\cite{DI}.
An equivalent derivation 
could also be given 
in the framework of {\it loop gravity\/} \cite{LQG}. 
But that would require a separate study beyond this
letter.

Summarizing we have seen that the Einstein-Hilbert action 
governs not only the classical physics 
of gravitational fields
but also, via the fluctuations of its line- and surface-like singularites, 
the quantum physics of dark matter. A string-like action, 
derived from it 
for the fluctuating surface-like singularites,
contains interacting spin-$2$ quanta that define a finite Quantum Gravity.

{~\\
Acknowledgment:
I am grateful to R. Kerr,
N. Hunter-Jones,
F. Linder, 
F. Nogueira,
A. Pelster, R. Ruffini, B. Schroer, and 
She-Sheng Xue
for useful comments.
}


\end{document}